\documentclass[10pt,twocolumn]{IEEEtran}
\usepackage{amsmath}
\usepackage{amssymb}
\usepackage{mathrsfs}
\usepackage{cite}
\usepackage{epsfig}
\usepackage{epsf}
\usepackage{theorem}
\usepackage{graphics}
\usepackage{subfigure}
\usepackage{url}
\usepackage{float}
\usepackage{multirow}
\usepackage{array}
\usepackage{hyperref}

\newenvironment{theorem}[2][Theorem]{\begin{trivlist}
\item[\hskip \labelsep {\bfseries #1}\hskip \labelsep {\bfseries #2}]}{\end{trivlist}}

\date{}

\begin{document}
  \title{\bf  On the Symmetric Gaussian Interference Channel with Partial Unidirectional Cooperation}
\author{
\authorblockN{\small Hossein Bagheri, Abolfazl
S. Motahari, and Amir K. Khandani}\\
\thanks{Financial support provided by Nortel and the corresponding matching funds by the Natural Sciences
and Engineering Research Council of Canada (NSERC), and Ontario
Centres of Excellence (OCE) are gratefully acknowledged.}
\thanks{The authors are affiliated with the Coding and Signal Transmission Laboratory,
Electrical and Computer Engineering Department, University of
Waterloo, Waterloo, ON, N2L 3G1, Canada, Tel: 519-884-8552, Fax:
519-888-4338, Emails:~ \{hbagheri, abolfazl,
khandani\}@cst.uwaterloo.ca.}} \maketitle
\begin{abstract}
A two-user symmetric Gaussian Interference Channel (IC) is considered
in which a noiseless unidirectional link connects one encoder to the other. Having a constant capacity, the additional link provides partial cooperation between the encoders. It is shown that the available cooperation can dramatically increase the sum-capacity of the channel. This fact is proved based on comparison of proposed lower and upper bounds on the sum-capacity. Partitioning the data into three independent messages, namely private, common, and cooperative ones, the transmission strategy used to obtain the lower bound enjoys a simple type of Han-Kobayashi scheme together with a cooperative communication scheme. A Genie-aided upper bound is developed which incorporates the capacity of the cooperative link. Other upper bounds are based on the sum-capacity of the Cognitive Radio Channel and cut-set bounds. For the strong interference regime, the achievablity scheme is simplified to employ common and/or cooperative messages but not the private one. Through a careful analysis it is shown that the gap between these bounds is at most one and two bits per real dimension for strong and weak interference regimes, respectively. Moreover, the Generalized Degrees-of-Freedom of the channel is characterized.
\end{abstract}
\section{Introduction}
Interference is one of the major limiting factors in achieving the total throughput of a network consisting of multiple non-cooperative transmitters intending to convey independent messages to their corresponding receivers through a common bandwidth.
The way interference is usually dealt with is either by treating it as noise or preventing it by associating different orthogonal dimensions, e.g. time or frequency division, to different users. Since interference has structure, it is possible for a receiver to decode some part of the interference and remove it from the received signal. This is, in fact, the coding scheme proposed by Han-Kobayashi (HK) for the two-user Gaussian IC \cite{HKIT81}. The two-user Gaussian IC provides a simple example showing that a single strategy against interference is not optimal. In fact, one needs to adjust the strategy according to the channel parameters \cite{Abolfazl_ISIT_08, Shang_ISIT_08, Annapureddy_ISIT_08}. However, a single suboptimal strategy can be proposed to achieve up to a single bit per user of the capacity region of the two-user Gaussian IC \cite{EtkinIT07}.

If the senders can cooperate, interference management can be done more effectively through cooperation. Cooperative links can be
either orthogonal (using out-of-band signalling) or non-orthogonal (using in-band signalling) to the shared medium \cite{SimeoneIT08}. In this work, unidirectional orthogonal cooperation is considered.

\textbf{\emph{Prior Works}}. Transmitter coordination over
orthogonal links is considered in different scenarios using
two-transmitter configurations as indicated in \cite{WillemsIT83,
NgIT071, MaricIT07, MaricET08,DevroyeIT06, WuIT07, JovicicIT06,
SimeoneIT08}. The capacity region of the Multiple Access Channel (MAC)
with bidirectional cooperation is derived in \cite{WillemsIT83},
where cooperation is referred to as \emph{conference}. Several achievable rate
regions are proposed for the IC with bidirectional transmitter and receiver
cooperation \cite{NgIT071}. The transmit
cooperation employs Dirty Paper Coding (DPC), whereas the receive cooperation uses
Wyner-Ziv
coding \cite{KramerIT05}. In the
transmit cooperation, the entire message of each transmitter is
decoded by the other cooperating transmitter, which apparently
limits the performance of the scheme to the capacity of the
cooperative link. The capacity regions of the compound MAC with
bidirectional transmitter cooperation and the IC with
Unidirectional Cooperation (ICUC) under certain strong interference conditions are obtained in \cite{MaricIT07}. These rate regions are shown to be closely
related to the capacity region of the Compound MAC with Common
Information (CMACCI). Furthermore, the inner and outer bounds for
the ICUC considered in \cite{MaricIT07} are given in \cite{MaricET08} for different interference
regimes. The above papers considered the ICUC in which the message sent by one of the encoders is \emph{completely} known to the other encoder. This setup is referred to as \emph{cognitive radio} in the
literature and is also considered in \cite{DevroyeIT06, WuIT07,
JovicicIT06}. More recently, the capacity region of the compound MAC
with both encoder and decoder unidirectional cooperation, and
physically degraded channels is derived in \cite{SimeoneIT08}. In a related problem, the sum-capacity of the ICUC channel with a different transmission model is characterized upto 9 bits \cite{Prabhakaran09}.

\textbf{\emph{Contribution}}.
For a Symmetric ICUC (SICUC) setup shown in Fig. 1 when $aP>1$, a simple
HK scheme in conjunction with cooperative communication is employed in the weak interference regime, \emph{i.e.}, $a\leq 1$.
In particular, each user partitions its data into three independent messages, namely private, common and
cooperative, and constructs independent codebooks for each of them. Each receiver, first jointly decodes the common and cooperative messages of both users and then decodes its own private message. Following \cite{EtkinIT07},
the power of the private message of each user is set
such that it is received at the level of the Gaussian noise at
the other receiver. This is to guarantee that the interference
caused by the private message has a small effect on the
other user. To prove that the scheme has negligible gap from the sum-capacity for \emph{all} transmit powers and channel gains, the transmit power should be properly allocated among the messages used by each user. Optimizing the
achievable sum-rate requires finding an optimum Power Allocation (PA)
over different codebooks employed by each user, which is mathematically involved.
For the weak interference regime, a simple power allocation strategy that works for both users and operates over the three codebooks is then proposed. Noting that the private codebook contributes in a small rate for the strong interference regime, \emph{i.e.}, $a> 1$, the scheme is subsequently modified to achieve a smaller gap and complexity by utilizing one or both of the common and cooperative codebooks but not
the private codebook. The achievable sum-rates are compared
to appropriate upper bounds to prove that the gap from the sum-capacity of the channel is one and two bits per real dimension, for the strong and weak interference regimes, respectively. The upper bounds are based on the cut-set bound and the sum-capacity of the Cognitive Radio Channel (CgRC), in which the message sent by one of the encoders is known to the other (cooperating) encoder.
A new upper bound is also proposed to incorporate the effect of the imperfect cooperative link, \emph{i.e.}, the case that only part of the message sent by one of the encoders is known to the other (cooperating) encoder \cite{Hossein_GDOF_IC09}. To obtain some asymptotic results, the
Generalized Degrees of Freedom (GDOF) of the channel is characterized. The GDOF represents the number of dimensions available for communication in
the presence of the interference when Signal-to-Noise Ratio $\text{SNR} \rightarrow \infty$. When SICUC is noise-limited ($aP\leq1)$, it is shown that treating interference as noise, and
not using the cooperative link, gives a sum-rate within a
single bit to the sum-capacity of the channel for all channel
parameters.

The rest of this paper is organized as follows: Section II presents the
system model. Sections III and IV, respectively, focus on the weak
and strong interference regimes. For each regime, the
corresponding achievable sum-rate, power allocation, upper bounds, and
the gap analysis are provided. Section V characterizes the GDOF of the channel. Section VI considers the noise-limited regime. Finally, section VII concludes the paper.

\textbf{\emph{Notation}}.
The sequence $\textbf{x}^{n}$ denotes $x_{1},\cdots,x_{n}$. $\bar{x}\!\triangleq\!1\!-\!x$. $\text{Prob}(\cdot)$ indicates the probability of an event, and $\mathbb{E}[X]$ represents the
expectation over the random variable $X$. All logarithms
are to the base 2 and all rates are expressed in bits per real
dimension. Finally, $C(P)\triangleq\frac{1}{2}\log(1+P)$.
\section{System Model} \label{sec: System Model}
In this work, a two-user symmetric Gaussian interference channel
with partial unidirectional cooperation, as depicted in Fig.
1, is considered. The model consists of two
transmitter-receiver pairs, in which each transmitter wishes to convey
its own data to its corresponding receiver. There exists a noiseless
cooperative link with capacity
$C_{12}$ from encoder $1$ to encoder $2$. It is assumed that all nodes
are equipped with a single antenna.
\begin{figure}[tb]
\centering
\includegraphics[width=0.25\textwidth]{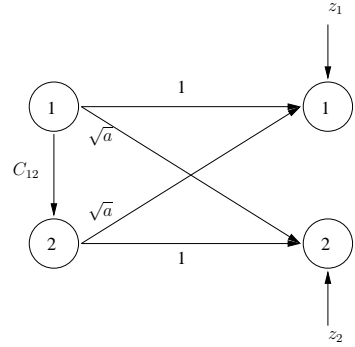}
\caption{The
symmetric interference channel with unidirectional transmitter cooperation.}
\end{figure} \label{fig: model}The input-output relationship for this channel in standard form is
expressed as \cite{SasonIT04}:
\begin{equation}
\begin{array}{rl}
  y_{1} &= x_{1}+\sqrt{a}\ x_{2}+z_{1}, \\
  y_{2} &= \sqrt{a}\ x_{1}+x_{2}+z_{2},
\end{array}
\end{equation}
where constant $a\!\geq\!0$ represents the gain of the interference
links. For $i\!\in\!\{1,2\}$, $z_{i}\!\sim\!\mathcal{N}(0,1)$. The average
power constraint of each transmitter is $P$, \emph{i.e.},
$\mathbb{E}[|x_{i}^{2}|]\leq P$.
The full Channel State Information (CSI) is assumed to be available at
the transmitters as well as receivers. Following \cite{EtkinIT07}, $\text{SNR}\!=\!P$ and Interference-to-Noise
Ratio $\text{INR}\!=\!aP$ are defined as the characterizing parameters of SICUC.

For a given block length $n$, encoder $i$
sends its own (random) message index $m_{i}$ from the index set
$\mathcal{M}_{i}\!=\!\{1,2,...,M_{i}=2^{nR_{i}}\}$ with rate $R_{i}$
[bits/channel use]. Each pair $(m_{1},m_{2})$ occurs with the same
probability $1/M_{1}M_{2}$. A one-step conference is made up of a communicating function $k$ which
maps the message index $m_{1}$ into $\textbf{q}^{n}$ with finite alphabet size
$|\mathcal{Q}|\!=\!2^{nC_{12}}$. The encoding function $f_{1}$ maps the
message index $m_{1}$ into a codeword $\textbf{x}_{1}^{n}$ chosen
from the codebook $\mathcal{C}_{1}$. The encoding function $f_{2}$ maps the
message index $m_{2}$ and what was obtained from the conference with
encoder $1$ into a codeword $\textbf{x}_{2}^{n}$ selected from
the codebook $\mathcal{C}_{2}$. Therefore:
\begin{equation}\label{eq: v_def}
\begin{array}{rl}
\textbf{q}^{n} &= k(m_{1}), \\
  \textbf{x}_{1}^{n} &= f_{1}(m_{1}), \\
  \textbf{x}_{2}^{n} &= f_{2}(m_{2},\textbf{q}^{n}).
  \end{array}
\end{equation}
The codewords in each codebook must satisfy the average power
constraint $\frac{1}{n}\sum_{t=1}^{n}|x_{i,t}|^2\leq P$ for $i\in\{1,2\}$.
 Each decoder uses a decoding function
$g_{i}(\textbf{y}_{i}^{n})$ to decode its
desired message index $m_{i}$ based on its received sequence. Let $\hat{m}_{i}$ be the output of the decoder. The
average probability of error for each decoder is:
$P_{e_{i}}\!=\!\mathbb{E}[\text{Prob}(\hat{m}_{i}\neq m_{i})]$.
A rate pair ($R_{1}$, $R_{2}$) is said to be achievable when there
exists an ($M_{1},M_{2},n,P_{e_{1}},P_{e_{2}}$)-code for the ICUC
consisting of two encoding functions $\{f_{1}, f_{2}\}$ and two
decoding functions $\{g_{1}, g_{2}\}$ such that for sufficiently
large $n$:
\begin{equation*}
\begin{array}{rl}
  R_{1} &\leq \frac{1}{n} \log(M_{1}),\\
  R_{2} &\leq \frac{1}{n} \log(M_{2}), \\
  P_{e} &\leq \epsilon.
  \end{array}
  \end{equation*}
In the above, $P_{e}\!=\!\max(P_{e_{1}}, P_{e_{2}})$ and $\epsilon\!\geq
\!0$ is a constant that can be chosen arbitrarily small. The capacity
region of the ICUC is the closure of the set of achievable rate pairs.
In this work, the characterization of the sum-capacity of the channel is considered. In addition, the interference-limited
regime is mainly investigated, \emph{i.e.}, $\text{INR}\!\geq\!1$ since otherwise
the system is noise limited and is not of much interest
\cite{EtkinIT07, Hossein_GDOF_IC09}. In fact, it will be shown later that treating interference as noise, and
not using the cooperative link, will give a sum-rate within a
single bit to the sum-capacity of the channel for all channel
parameters. Finally, for rate derivations in this paper, Gaussian
codebooks are employed.
\section{Weak Interference Regime ($a\leq 1$)}
\subsection{Achievable Sum-Rate}
A simple type of the Han-Kobayashi scheme is used in
\cite{EtkinIT07} for the symmetric IC, in which the codebook of transmitter $i$ is
composed of private and common codebooks denoted by
$\mathcal{C}_{i}^{u}$ and $\mathcal{C}_{i}^{w}$, respectively. The
transmitted signal is $x_{i}=\sqrt{P_{u}}u_{i}+\sqrt{P_w}w_{i}$,
and it is assumed that $\mathbb{E}[|u_{i}|^{2}]=\mathbb{E}[|w_{i}|^{2}]=1$.
The power associated with the codebooks are $P_{u}$ and $P_{w}$,
such that $P_{u}+P_{w}=P$. First, the common messages are decoded at both
decoders and their effect is subtracted from the received signal. Then, the private messages are decoded at
their corresponding receiver while treating the private signal of the other user as noise. One of the main steps towards getting a small gap is
that the power of the interfering private message is set to be at the noise
level, \emph{i.e.}, $aP_{u}=1$ \cite{EtkinIT07}.

In this work, because the existence of the cooperative link, encoder $2$ can
relay the transmitter 1's information to help the
other receiver \cite{cover_book}. Therefore, it is natural to have three codebooks
at encoder $i$, namely private codebook $\mathcal{C}_{i}^{u}$
with codewords $u_{i}$, common codebook $\mathcal{C}_{i}^{w}$ with
codewords $w_{i}$ , and cooperative codebook $\mathcal{C}_{i}^{v}$
with codewords $v$. The cooperative message is decoded error
free at transmitter 2 and relayed to both receivers. The input
to the channel can be written as:
\begin{IEEEeqnarray}{rl}
X_{i} &= \sqrt{P_{u}}u_{i}+\sqrt{P_{w_{i}}}w_{i}+\sqrt{P_{v_{i}}}v.
\end{IEEEeqnarray}
The average power constraint for transmitter $i$ is $P_{u}+P_{w_{i}}+P_{v_{i}} = P$.
The following power allocation is used:
\begin{equation}\label{eq: power}
\begin{array}{rl}
 P_{u} &= \frac{1}{a}, \\
  P_{v_{i}} &= \gamma_{i}(P-\frac{1}{a}), \\
  P_{w_{i}} &=\bar{\gamma}_{i}(P-\frac{1}{a}),
    \end{array}
\end{equation}
where $0\!\leq\!\gamma_{i}\!\leq\!1$ is the corresponding power allocation
parameter for transmitter $i$. The received power of $v$ at
decoder 1 is: $P_{V} = (\sqrt{P_{v_{1}}}+\sqrt{a
  P_{v_{2}}})^{2}$.

Because of the symmetries in the
problem (in terms of the channel gains), and the fact that the sum-rate is the objective of this paper, we set $\gamma_{1}\!=\gamma_{2}\!=\gamma$. It is also assumed that $v$ is decoded at both decoders.
Therefore, the achievable
rate pairs associated with the HK scheme are in fact the intersection of
the capacity regions of two MACs: each composed of four virtual
users, \emph{i.e.}, $\{u_{1},w_{1},w_{2},v\}$ for MAC$_{1}$, and
$\{u_{2},w_{1},w_{2},v\}$ for MAC$_{2}$. Among all rate assignments
and decoding orders for the MACs, the following scheme is used. It will be shown later that the
scheme achieves within two bits of the
associated upper bound.
The strategy is along the same line as the approach of
\cite{EtkinIT07} for the IC. The common codewords $w_{1}, w_{2}$, and the cooperative codeword $v$ are jointly decoded at both receivers, treating the interfering private signals as noise. Finally, each receiver decodes its private codeword\footnote{It is remarked that encoder 2 could perform the Gel'fand-Pinsker binning \cite{GelfandPr80} in order to precode its own message against the known interference. This, in general, could increase the rate at receiver 2, especially in the weak interference regime. In this paper, however we show that one can achieve a sum-rate within two bits of the sum-capacity of SICUC without binning. Employing the binning technique will be considered in future works.}.
Therefore, the capacity region of four-user MACs are projected
onto three-dimensional subspaces. As a result, a compound
MAC with three virtual users ($w_{1},w_{2},v$) is obtained. The capacity region
of MAC$_{i}$ $i\in\{1,2\}$ is:
\begin{IEEEeqnarray}{rl}
  \label{eq: R_v}R_{v} &\leq \min\{C(\frac{P_{V}}{P_{u}+2}),C_{12}\}, \\
  \label{eq: R_w1}R_{w_{i}} &\leq C(\frac{P_{w}}{P_{u}+2}), \\
  \label{eq: R_w2}R_{w_{j}} &\leq C(\frac{aP_{w}}{P_{u}+2}), \\
  \label{eq: R_w1_w2}R_{w_{i}}+R_{w_{j}} &\leq C(\frac{P_{w}(1+a)}{P_{u}+2}), \\
  \label{eq: R_w1_v}R_{w_{i}}+R_{v} &\leq C(\frac{P_{w}+P_{V}}{P_{u}+2}), \\
  \label{eq: R_w2_v}R_{w_{j}}+R_{v} &\leq C(\frac{aP_{w}+P_{V}}{P_{u}+2}), \\
 \label{eq: R_w1_w2_v}R_{w_{i}}+R_{w_{j}}+R_{v} &\leq
 C(\frac{P_{w}+aP_{w}+P_{V}}{P_{u}+2}),
 \end{IEEEeqnarray}
where $j\in\{1,2\}$, and $j\neq i$. Also, $P_{w_{1}}\!=\!P_{w_{2}}\!\triangleq\!P_{w}$, $P_{v_{1}}\!=\!P_{v_{2}}\!\triangleq\!P_{v}$, and
\begin{equation}\label{eq: PV}
    P_{V}=(1+\sqrt{a})^{2}P_{v}.
\end{equation}
The factor 2 in the denominators
comes from the fact that the power of the interfering private signal
has been set to unity. It is also assumed that:
$R_{w_{1}}\!=\!R_{w_{2}}\!\triangleq\!R_{w}$. Since the sum-rate is $2(R_{u}\!+\!R_{w})\!+\!R_{v}$, in the above,
$2R_{w}+R_{v}$ is of particular interest. There exist three ways to construct $2R_{w}+R_{v}$ from inequalities (\ref{eq: R_v}-\ref{eq: R_w1_w2_v}):
\begin{IEEEeqnarray}{rl}
  \label{eq: SR1}2R_{w}+R_{v}&\leq C(\frac{P_{w}(1+a)+P_{V}}{P_{u}+2})\triangleq R_{\mathcal{B}_{1}},\\
    \label{eq: SR2}2R_{w}+R_{v}&\leq
  2\tilde{R}_{w}+\min\{C(\frac{P_{V}}{P_{u}+2}),C_{12}\}\triangleq R_{\mathcal{B}_{2}},\\
\label{eq: SR3}2R_{w}+R_{v}&\leq
  C(\frac{aP_{w}+P_{V}}{P_{u}+2})+\tilde{R}_{w}\triangleq R_{\mathcal{B}_{3}} ,
\end{IEEEeqnarray}
where
\begin{equation}
    \tilde{R}_{w}\triangleq
    \min\{C(\frac{aP_{w}}{P_{u}+2}),\frac{1}{2}C(\frac{P_{w}(1+a)}{P_{u}+2})\},
\end{equation}
and (\ref{eq: SR1}-\ref{eq: SR3}) come from inequalities (\ref{eq: R_w1_w2_v}), (\ref{eq: R_v}-\ref{eq: R_w2}), and (\ref{eq: R_w1}, \ref{eq: R_w2}, \ref{eq: R_w1_v}, \ref{eq: R_w2_v}), respectively.
The achievable sum-rate is:
\begin{IEEEeqnarray}{rl}
  \label{eq: sum-rate} R_{\text{sum}}&=\max_{0 \leq \gamma\leq 1}\big\{2C(\frac{P_{u}}{2})+\min_{i\in\{1,2,3\}}\{R_{\mathcal{B}_{i}}\}\big\}.
\end{IEEEeqnarray}
Optimizing the achievable sum-rate requires finding optimum power
allocation parameter $\gamma$, which is mathematically involved.
 In the following, a simple power allocation
strategy is provided, and is later shown to be within two bits
of the related upper bound.
\subsection{Power Allocation Policies and the Associated Sum-Rates}
\subsubsection{Universal PA}
As will be shown later, the following power allocation guarantees a small gap form the sum-capacity for all $a\leq 1$ and hence called \emph{universal}.
 We set $P_{w}\!=\!P_{V}$, which together with
$P_{u}\!=\!\frac{1}{a}$ give:
\begin{equation}\label{eq: PA}
P_{w}=\frac{P-\frac{1}{a}}{1+\frac{1}{(1+\sqrt{a})^{2}}}.
\end{equation}
The corresponding sum-rate is $R\!+\!2C(\frac{P_{u}}{2})$, where
\begin{equation}\label{eq: R_min}
    R\!\triangleq\!\min_{i\in\{1,\cdots,5\}}\!\{R_{i}\},
\end{equation}
and
\begin{IEEEeqnarray}{rl}\label{eqnarr: R1_5 }
  R_{1} &= C(\frac{(2+a)P_{w}}{P_{u}+2}), \nonumber\\
  R_{2} &= 2C(\frac{aP_{w}}{P_{u}+2})+C_{12}, \nonumber\\
  R_{3} &= C(\frac{(1+a)P_{w}}{P_{u}+2})+C_{12}, \nonumber\\
  R_{4} &= C(\frac{(1+a)P_{w}}{P_{u}+2})+C(\frac{aP_{w}}{P_{u}+2}), \nonumber\\
  R_{5} &= \frac{3}{2}C(\frac{(1+a)P_{w}}{P_{u}+2})
\end{IEEEeqnarray}
are obtained from inequalities (\ref{eq: SR1}-\ref{eq: SR3}), Eq. (\ref{eq: PA}). Note that the following redundant inequalities are eliminated:
\begin{IEEEeqnarray*}{rl}
  R_{6} &= 2C(\frac{aP_{w}}{P_{u}+2})+C(\frac{P_{w}}{P_{u}+2}),\\
  R_{7} &= C(\frac{(1+a)P_{w}}{P_{u}+2})+C(\frac{P_{w}}{P_{u}+2}).
\end{IEEEeqnarray*}
It is remarked that this allocation achieves the optimal GDOF of the channel as will be shown later and therefore can be among the achievable schemes that have a small gap from the sum-capacity for all channel parameters\footnote{In fact, if an achievable scheme fails to achieve the GDOF of the channel it cannot have a small gap from the sum-capacity for all channel parameters.}. This allocation also simplifies the gap analysis.
\subsubsection{Full Cooperation PA}
It is also interesting to consider the case that all the remaining $P-\frac{1}{a}$ power is devoted to cooperation or equivalently $P_{w}\!=\!0$,
\emph{i.e.}, $\gamma\!=\!1$ in (\ref{eq: power}). In this case, the
achievable sum-rate becomes:
\begin{equation}\label{eq: PA_Rate}
    R_{\text{FC}}=\min\{C_{12},C(\frac{(1+\sqrt{a})^{2}(aP-1)}{2a+1})\}+2C(\frac{P_{u}}{2}).
\end{equation}
It will be shown that this power allocation provides a smaller gap compared to the universal PA in some regions. However, it does not achieve the GDOF \emph{universally} \cite{Hossein_GDOF_IC09}, and hence cannot provide a small gap for all channel parameters.
\subsection{Upper Bounds}
Two upper bounds are used for this regime. The first one is an
enlarged version of the sum-capacity of the Cognitive Radio Channel
(CgRC) with weak interference \cite{WuIT07}. The
sum-capacity of the CgRC for $a\leq 1$, denoted by $C_{\text{sum}}^{w}$, is \cite{WuIT07}:
\begin{align*}
&C_{\text{sum}}^{w}=\\
    &\frac{1}{2}\max_{0 \leq \eta \leq 1}\bigg[\log(1+aP+2\sqrt{\bar{\eta}a} P +
    P) + \log(\frac{1+\eta P}{1+\eta aP})\bigg].
   \end{align*}
It is noted that the first and the second terms in the above upper bound
are monotonically decreasing and increasing functions of $\eta$,
respectively. The above upper bound is enlarged by setting
$\bar{\eta}=1$ and $\eta=1$ in the first and second terms,
respectively. Therefore, the following upper bound serves as the first upper bound for the region:
\begin{equation}\label{eq: ub_1}
    R_{\text{ub}}^{(1)}=\frac{1}{2}\log\big(1+(1+\sqrt{a})^{2}P\big) + \frac{1}{2}\log(\frac{1+ P}{1+ aP}).
\end{equation}

For the IC ($C_{12}=0$), Etkin, \emph{et.al} \cite{EtkinIT07} provide
a new upper bound by selecting an appropriate genie. In the following, the effect of the cooperative link is incorporated in the genie-aided upper
bound. The genie signal $s_{i}, \ i={1,2}$ is provided to
receiver $i$:
\begin{equation}\label{eq: genie}
\begin{array}{rl}
  s_{1} &= \sqrt{a}x_{1}+z_{2}, \\
  s_{2} &= \sqrt{a}x_{2}+z_{1}.
  \end{array}
\end{equation}
The upper bound on the sum-capacity of the genie-aided channel is:
\begin{IEEEeqnarray}{rl}
  nR_{\text{sum}}&\leq I(m_{1};\textbf{y}_{1}^{n},\textbf{s}_{1}^{n},\textbf{q}^{n})+I(m_{2};\textbf{y}_{2}^{n},\textbf{s}_{2}^{n},\textbf{q}^{n})+
  n\epsilon_{n}\nonumber\\
  {}&\stackrel{\text{(a)}}{=}I(m_{1};\textbf{q}^{n})
  +I(m_{1};\textbf{y}_{1}^{n},\textbf{s}_{1}^{n}|\textbf{q}^{n})+I(m_{2};\textbf{q}^{n})
  \nonumber\\
  {}&{}+I(m_{2};\textbf{y}_{2}^{n},\textbf{s}_{2}^{n}|\textbf{q}^{n})+n\epsilon_{n}\nonumber\\
  {}&\stackrel{\text{(b)}}{=}h(\textbf{q}^{n})+I(m_{1};\textbf{y}_{1}^{n},\textbf{s}_{1}^{n}|\textbf{q}^{n})+\nonumber\\
  {}& I(m_{2};\textbf{y}_{2}^{n},\textbf{s}_{2}^{n}|\textbf{q}^{n})+n\epsilon_{n}\nonumber\\
  {}&\leq\sum_{t=1}^{n} h(q_{t}) + I(m_{1};\textbf{s}_{1}^{n}|\textbf{q}^{n})+I(m_{1};\textbf{y}_{1}^{n}|\textbf{s}_{1}^{n},\textbf{q}^{n})\nonumber\\
  {}&+I(m_{2};\textbf{s}_{2}^{n}|\textbf{q}^{n})+I(m_{2};\textbf{y}_{2}^{n}|\textbf{s}_{2}^{n},\textbf{q}^{n})+n\epsilon_{n}\nonumber\\
  {}&=nC_{12}+ h(\textbf{s}_{1}^{n}|\textbf{q}^{n})-h(\textbf{s}_{1}^{n}|m_{1},\textbf{q}^{n})+\nonumber\\
  {}&I(m_{1};\textbf{y}_{1}^{n}|\textbf{s}_{1}^{n},\textbf{q}^{n})+ h(\textbf{s}_{2}^{n}|\textbf{q}^{n})-h(\textbf{s}_{2}^{n}|m_{2},\textbf{q}^{n})+\nonumber\\
  {}&I(m_{2};\textbf{y}_{2}^{n}|\textbf{s}_{2}^{n},\textbf{q}^{n})+n\epsilon_{n}\nonumber\\
  {}&\stackrel{\text{(c)}}{=} nC_{12}+h(\textbf{s}_{1}^{n}|\textbf{q}^{n})-h(\textbf{s}_{1}^{n}|\textbf{x}_{1}^{n})+{}\nonumber\\
  {}& h(\textbf{y}_{1}^{n}|\textbf{s}_{1}^{n},\textbf{q}^{n})-h(\textbf{y}_{1}^{n}|\textbf{s}_{1}^{n},\textbf{q}^{n},m_{1},\textbf{x}_{1}^{n}(m_{1}))+{}\nonumber\\
  {}&h(\textbf{s}_{2}^{n}|\textbf{q}^{n})-h(\textbf{s}_{2}^{n}|\textbf{x}_{2}^{n})+h(\textbf{y}_{2}^{n}|\textbf{s}_{2}^{n},\textbf{q}^{n}){}\nonumber\\
  {}& -h(\textbf{y}_{2}^{n}|\textbf{s}_{2}^{n},\textbf{q}^{n},m_{2},\textbf{x}_{2}^{n}(m_{2},\textbf{q}^{n}))+n\epsilon_{n}\nonumber\\
  {}&\stackrel{\text{(d)}}{\leq} nC_{12}+h(\textbf{s}_{1}^{n}|\textbf{q}^{n})-h(\textbf{z}_{2}^{n})+h(\textbf{y}_{1}^{n}|\textbf{s}_{1}^{n}){}\nonumber\\
  {}& -h(\textbf{y}_{1}^{n}|\textbf{s}_{1}^{n},\textbf{q}^{n},m_{1},\textbf{x}_{1}^{n}(m_{1}))+h(\textbf{s}_{2}^{n}|\textbf{q}^{n}){}\nonumber\\
  {}&-h(\textbf{z}_{1}^{n})+h(\textbf{y}_{2}^{n}|\textbf{s}_{2}^{n}){}\nonumber\\
  {}& -h(\textbf{y}_{2}^{n}|\textbf{s}_{2}^{n},\textbf{q}^{n},m_{2},\textbf{x}_{2}^{n}(m_{2},\textbf{q}^{n}))+n\epsilon_{n}\nonumber\\
  {}&\stackrel{\text{(e)}}{=} nC_{12}-h(\textbf{z}_{1}^{n})-h(\textbf{z}_{2}^{n})+h(\textbf{y}_{1}^{n}|\textbf{s}_{1}^{n})+\nonumber\\
  {}& h(\textbf{y}_{2}^{n}|\textbf{s}_{2}^{n})+n\epsilon_{n},
\end{IEEEeqnarray}
where (a), (b), and (c) respectively, follow from the chain
rule of mutual information \cite{cover_book}, Eq. (\ref{eq: v_def}),
and Markovity of $m_{1}\rightarrow \textbf{x}_{1}^{n} \rightarrow
\textbf{s}_{1}^{n}$ and $(\textbf{q}^{n}, m_{2})\rightarrow
\textbf{x}_{2}^{n} \rightarrow \textbf{s}_{2}^{n}$. (d) comes from
Eq. (\ref{eq: genie}) and the fact that removing the condition
$\textbf{q}^{n}$ does not decrease the entropy. (e) is valid because $\textbf{y}_{2}^{n}$ and $\textbf{s}_{2}^{n}$ are independent conditionally on $(\textbf{x}_{2}^{n},\textbf{q}^{n})$. The same fact is applied to $\textbf{y}_{1}^{n}$ and $\textbf{s}_{1}^{n}$ conditionally on $(\textbf{x}_{1}^{n},\textbf{q}^{n})$. Following the same
reasoning as \cite{EtkinIT07} for
$h(\textbf{y}_{i}^{n}|\textbf{s}_{i}^{n})$, $i\in\{1,2\}$, we get
the following upper bound:
\begin{equation}\label{eq: ub_2}
    R_{\text{ub}}^{(2)}=C_{12}+ \log\Big(\frac{P+ (aP+1)^{2}}{aP+1}\Big).
\end{equation}
We remark that the above upper bound is the same as the upper bound
derived in \cite{EtkinIT07}, except for the additional term
$C_{12}$.

\subsection{Gap Analysis}
\begin{theorem}{1}
For $a\leq 1$, the achievable sum-rate associated
with the universal power allocation policy of (\ref{eq: PA}) is
within two bits of the sum-capacity of the channel.
\end{theorem}
\begin{IEEEproof}
See Appendix \ref{appendix: Proof_1}.
\end{IEEEproof}
In the next theorem, the maximum gap is shown to be smaller than 2 bits using full cooperation for some cases.
\begin{theorem}{2}
The achievable sum-rate corresponding to the
full cooperation power allocation policy (Eq. (\ref{eq: PA_Rate})) is within 1 and 1.5 bits of the
sum-capacity of the channel for the following cases, respectively.

\begin{enumerate}
  \item CASE I: $C_{12}\!\leq\!C(\frac{b(aP-1)}{2a+1})$ and $a^{3}P^{2}\!\leq\!a\!+\!1$\\
  \item CASE II: $C_{12}\!\geq\!C(\frac{b(aP-1)}{2a+1})$,\\
\end{enumerate}
where
\begin{equation}\label{eq: b}
    b\triangleq(1+\sqrt{a})^{2},
\end{equation}
\begin{IEEEproof}For each case, the gap $\Delta_{\textrm{FC}}$ is calculated:
\begin{enumerate}

  \item CASE I:
  The achievable sum-rate is:
  \begin{equation*}
    R_{\text{sum}}\!=\!C_{12}+2C(\frac{1}{2a}).
  \end{equation*}
  The gap from the second upper bound, \emph{i.e.}, Eq. (\ref{eq: ub_2}) is:  \begin{equation*}
    \Delta_{\textrm{FC},\text{I}}=1+\log\big(1+\frac{a^{3}P^{2}-a-1}{(aP+1)(2a+1)}\big)\leq 1.
  \end{equation*}

    \item CASE II: In this case, the achievable sum-rate
    \begin{equation*}
        R_{\text{sum}}\!=\!\frac{1}{2}\big[\log\big(\frac{2a+1+b(aP-1)}{a}\big)\!+\!\log(\frac{2a+1}{a})\big]\!-\!1
    \end{equation*}
   %
is compared to the upper bound of Eq. (\ref{eq: ub_1}):
\begin{IEEEeqnarray*}{ll}
       \Delta_{\textrm{FC},\text{II}}&=1+\frac{1}{2}\log(\frac{a+aP}{(2a+1)(1+aP)})+\\
       {}&\quad+C(\frac{2\sqrt{a}}{2a+1+b(aP-1)})\\
       {}&\leq1+C(\frac{2\sqrt{a}}{a+1})\\
       {}&\leq\frac{3}{2}.
\end{IEEEeqnarray*}
\end{enumerate}
\end{IEEEproof}
\end{theorem}

\section{Strong Interference Regime ($a\!>\!1$)}
\subsection{Achievable Sum-Rate}
 It is remarked that for $a\!\geq\!1$, there is little benefit using
the private codebooks because: $\log(1+\frac{P_{u}}{2})=\log(1+\frac{1}{2a})\leq 1$.
%
Therefore, we can modify the achievablity scheme to have only
$w_{1},w_{2},$ and $v$. Here, we propose
simple PA policies for
the following cases and prove that the gap between the corresponding achievable rates and sum-capacity is less than 1 bit for each scenario.
For $a\geq1$, three cases may happen:
\begin{enumerate}
  \item $P\!\leq\!1\!\leq\!a$
  \item $1\!\leq\!a\!\leq\!P$
  \item $1\!\leq\!P\!\leq\!a$
\end{enumerate}

For the first case, user 2 can at most get $C(P)\leq 1$. Since $a>1$ we can simply use all the available power for cooperation \emph{i.e.}, $P_{w}\!=\!P_{u}\!=\!0$.

The common codebooks play an important role for the last two cases in the IC without the cooperative link. As IC is a special case of ICUC with $C_{12}=0$, we give most of the available power to common codewords in these cases.
\subsubsection{$P\!\leq\!1\!\leq\!a$}
\vspace{5pt}
In this case, we let $P_{w}\!=\!P_{u}\!=\!0$ and $P_{V}\!=\!bP$. The achievable sum-rate becomes:
\begin{equation}\label{eq: rate_Pleq1}
    R_{\textrm{sum}}=\min\{C_{12},C(bP)\}.
\end{equation}

\subsubsection{$1\!\leq\!a\!\leq\!P$}
\vspace{5pt}
In this case, let $P_{w}=P$ and $P_{u}=P_{V}=0$. It is straightforward to show that
 $C\big((1+a)P_{w}\big)\leq
2C(P_{w})$. Therefore, the achievable sum-rate becomes:
\begin{equation}\label{eq: rate_SIR}
    R_{\textrm{sum}}=C\big((1+a)P\big).
\end{equation}

\subsubsection{$1\!\leq\!P\!\leq\!a$}
\vspace{5pt}
In this scenario, it is suggested that receiver 1 jointly decodes
$w_{2}, v$ first and then $w_{1}$. Similarly, receiver 2 jointly
decodes $w_{1}, v$ first and then $w_{2}$. The following power
allocation is used:
\begin{equation}
\begin{array}{rl}
  P_{w} &= P-1, \\
  P_{V} &= b.
  \end{array}
\end{equation}
The rate constraints are:
\begin{IEEEeqnarray}{rl}
  R_{v} &\leq \min\{C_{12}, C(\frac{P_{V}}{P_{w}+1})\}, \nonumber\\
  R_{w} &\leq \min\{C(P_{w}),C(\frac{aP_{w}}{P_{w}+1})\}, \nonumber\\
  R_{v}+R_{w} &\leq C(\frac{P_{V}+aP_{w}}{P_{w}+1}).\nonumber
\end{IEEEeqnarray}
Since $a\geq P$, the power allocation leads to:
\begin{equation*}
    R_{w}=\min\{C(P-1),C(\frac{a(P-1)}{P})\}=C(P-1).
\end{equation*}
Therefore, the achievable sum-rate becomes:
\begin{equation}\label{eq: ach_rate_2_alpha}
\begin{array}{rl}
    R_{\textrm{sum}}&=
    \min\Big\{\min\big\{C_{12},
    C(\frac{b}{P})\big\}+\\
    &{}+C(P-1),C(\frac{aP+1+2\sqrt{a}}{P})\Big\}+C(P-1).
    \end{array}
\end{equation}

\subsection{Upper Bounds}

Using the cut-set analysis \cite{cover_book}, transmitter 1 can send
at most up to $C_{12}+C(P)$ bits per channel use. It is clear that
transmitter 2 can provide at most $C(P)$ bits per channel use to
its corresponding receiver. Another upper bound is the sum-capacity
of the CgRC for the specific strong interference regime reported in
\cite{MaricIT07}. It is easy to verify that the SICUC with
$a\!\geq\!1$ satisfies the strong interference conditions of
\cite{MaricIT07}. Therefore, the sum-rate is upper bounded by:
\begin{equation}\label{eq: ub_3}
    R_{\text{ub}}^{(3)}=\min\left\{C_{12}+2C(P),C(bP)\right\}.
\end{equation}

\subsection{Gap Analysis}

\subsubsection{$P\!\leq\!1\!\leq\!a$}
\vspace{5pt}
 In this case, if $C_{12}\!\leq\!C(bP)$, the achievable sum-rate of Eq. (\ref{eq: rate_Pleq1}) is compared to the first term of the upper bound of  Eq. (\ref{eq: ub_3}), otherwise it is compared to the second term of Eq. (\ref{eq: ub_3}).
 Therefore the gap $\Delta\!\leq\!2C(P)\!\leq\!1$.

\subsubsection{$1\leq\!a\!\leq P$}
\vspace{5pt}
In this case, the difference ($\Delta$) between the second term in the
upper bound of Eq. (\ref{eq: ub_3}) and the achievable sum-rate of
Eq. (\ref{eq: rate_SIR}) is calculated as follows:
\begin{IEEEeqnarray}{rl}
  \Delta &= \frac{1}{2} \log\big(1+\frac{2\sqrt{a}P}{1+aP+P}\big) \nonumber\\
  {} &\leq \frac{1}{2} \log\big(1+\frac{aP+P}{1+aP+P}\big) \nonumber\\
  {} &\leq \frac{1}{2}.\nonumber
\end{IEEEeqnarray}

\subsubsection{$1\leq P\leq a$}
\vspace{5pt}
Because of Eq. (\ref{eq: ach_rate_2_alpha}) two cases can occur:
\begin{itemize}
  \item $C_{12}> C\big(\frac{b}{P}\big)$:
  For this condition, the achievable sum-rate of Eq. (\ref{eq: ach_rate_2_alpha}) is:
    \begin{IEEEeqnarray*}{rl}
    R_{\textrm{sum}}&=\frac{1}{2}\min\big\{\log(P^2+bP),\log(P+aP+1+2\sqrt{a})\big\}\\
    &=\frac{1}{2}\log(P+aP+1+2\sqrt{a}).
   \end{IEEEeqnarray*}

Then the gap with respect to Eq. (\ref{eq: ub_3}) becomes:
\begin{IEEEeqnarray*}{rl}
  \Delta &= \frac{1}{2}\log(1+\frac{2\sqrt{a}(P-1)}{2\sqrt{a}+1+aP+P}) \\
  {}&\leq \frac{1}{2}\log(1+\frac{2\sqrt{a}P}{1+aP+P}) \\
  {}&\leq \frac{1}{2}\log(1+\frac{aP+P}{1+aP+P}) \\
  {} &\leq \frac{1}{2}.
\end{IEEEeqnarray*}
%
  \item $C_{12}\leq C\big(\frac{b}{P}\big)$:
  In this case, the gap between the following achievable sum-rate:
  \begin{IEEEeqnarray*}{rl}
  R_{\textrm{sum}}&=\min\left\{C_{12}+\log(P), \frac{1}{2}\log(1+P+aP+2\sqrt{a})\right\}
  \end{IEEEeqnarray*}
and the upper bound of Eq. (\ref{eq: ub_3}) is:
\begin{equation*}
\begin{array}{rl}
    \Delta&\leq\max\left\{\log(1+\frac{1}{P}),\frac{1}{2}\log(1+\frac{2\sqrt{a}(P-1)}{1+P+aP+2\sqrt{a}})\right\}\\
    {}&\leq1.
    \end{array}
\end{equation*}
\end{itemize}

\subsection{Sum-Capacity for Sufficiently Large $C_{12}$}

In \cite{Hossein_GDOF_IC09}, we simply showed that if
\begin{equation}\label{eq: capacity condition}
    C_{12} \geq C\big(bP\big),
\end{equation}
by setting $P_{v_{1}}=P_{v_{2}}=P$, we can achieve the second term
of Eq. (\ref{eq: ub_3}) and hence, the sum-capacity of the channel.
\section{GDOF Analysis}
In this section, the sum-capacity behavior is considered in the high SNR regime by characterizing the GDOF of the channel.

It is known that the interference can reduce the available degrees
of freedom for data communication \cite{EtkinIT07}. To understand this effect, we define the GDOF as below:
\begin{equation}\label{eq: GDOF}
    d(\alpha, \beta)=\lim_{\text{SNR}\rightarrow
    \infty}{\frac{R_{\text{sum}}(\text{SNR}, \alpha, \beta)}{C(\text{SNR})}},
\end{equation}
where
\begin{equation}\label{eq: alpha_def}
    \alpha\triangleq\frac{\log(\text{INR})}{\log(\text{SNR})},
\end{equation}
and
$\beta\geq 0$ is the multiplexing gain of the cooperative link, \emph{i.e.},
\begin{equation}\label{eq: C_12}
    C_{12}\triangleq \beta C(P).
\end{equation}
Here, the GDOF of the proposed scheme with rate constraints (\ref{eq: SR1}-\ref{eq: SR3}) is derived.

 Using the power allocation (\ref{eq: power}) and Eq. (\ref{eq: alpha_def}), $a$ can be written as $a=P^{\alpha-1}$. As mentioned earlier, optimizing the achievable sum-rate of Eq. (\ref{eq: sum-rate}) requires finding optimum power
allocation parameter $\gamma$, which is mathematically involved.
To characterize the GDOF, the sum-rate from Eq. (\ref{eq: sum-rate}) for each region
$\mathcal{B}_{i}$ is calculated assuming a suboptimal fixed $\gamma$, for $P\rightarrow\infty$. Then the minimum of the sum-rates is considered for GDOF analysis. It is clear that this assumption on $\gamma$ provides an achievable GDOF that might be smaller than the GDOF obtained using the optimal $\gamma$. However, by deriving the GDOF associated with the considered upper bounds, it is easy to see that the assumption does not entail any GDOF loss. In fact the GDOF associated with the proposed scheme is the optimal GDOF for the SICUC. Theorem 3 provides the GDOF of the SICUC:

\begin{theorem}{3}
The optimal GDOF for the SICUC setup is:
\begin{equation}\label{eq: DoF}
d(\alpha, \beta) = \left\{ \begin{array}{ll}
2 - 2 \alpha + \min\big\{\alpha, \beta\big\}, & \alpha<\frac{1}{2}\\
\min\big\{2-\alpha, 2 \alpha + \min \{\alpha, \beta\} \big\}, & \frac{1}{2}\leq\alpha<\frac{2}{3}\\
2- \alpha, &
\frac{2}{3}\leq\alpha<1\\
\alpha, &
1\leq\alpha<2\\
\min\big\{2 + \beta, \alpha \big\}. & 2\leq\alpha
\end{array} \right.
\end{equation}
 \end{theorem}
 \begin{IEEEproof}
 The derivation involves straightforward computation, and hence is omitted.
 \end{IEEEproof}
It is remarked that if $\beta=0$, the same GDOF for the symmetric IC
derived in \cite{EtkinIT07} will be obtained.

\begin{figure}[tb]
\centering
\includegraphics[width=0.37\textwidth]{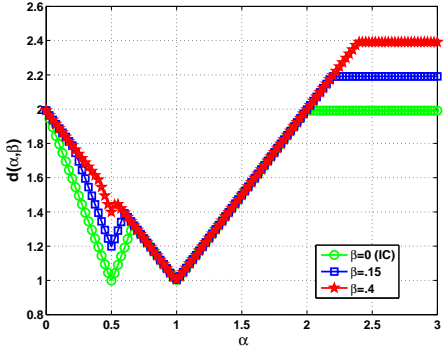}
\caption{The effect of partial cooperation on the GDOF of the SICUC for
$\beta<\frac{1}{2}$.} \label{fig: curve1}
\end{figure}

\begin{figure}[tb]
\centering
\includegraphics[width=0.37\textwidth]{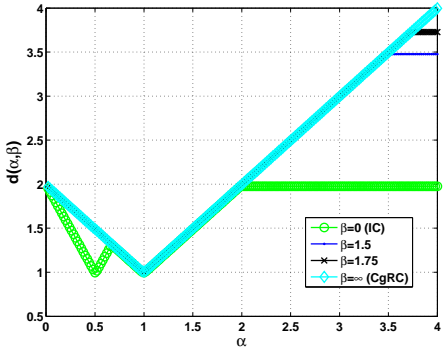}
\caption{The effect of partial cooperation on the GDOF of the SICUC for
$\frac{1}{2}\leq\beta$. $\beta=0$ is also considered for the purpose of comparison.}
\label{fig: curve2}
\end{figure}

Figures \ref{fig: curve1} and \ref{fig: curve2} show the GDOF as a
function of $\alpha$ and $\beta$. Fig. \ref{fig: curve1} gives the
GDOF for $\beta \leq \frac{1}{2}$. Compared to the GDOF of the IC,
one can realize that increasing the amount of cooperation
(corresponding to the value of $\beta$) is beneficial in terms of
the GDOF for the ranges $\alpha<\frac{2}{3}$ and $\alpha>2$. For
instance, it increases the GDOF from $2-2\alpha$ to $2-\alpha$ for
$\alpha\leq \frac{1}{2}$ and $\alpha\leq \beta$ compared to the IC
case. Fig. \ref{fig: curve2} demonstrates the GDOF for
$\frac{1}{2}\leq\beta$. Here, increasing the amount of cooperation
is not useful for $\alpha\leq 2$. For the purpose of comparison, the
GDOF of the two known extreme cases of $\beta=0$ \emph{i.e.}, the IC
\cite{EtkinIT07} and $\beta=\infty$, \emph{i.e.}, the CgRC \cite{MaricIT07}
are plotted as well.

\section{Signalling for $aP\leq 1$}
The interference-limited regime, \emph{i.e.}, $aP\geq 1$ has so far been considered in this paper. In this section, the sum-rate analysis is done for the case that noise is the major performance-limiting factor in the communication system shown in Fig. 1.
\begin{theorem}{4}
 If $aP\!\leq\!1$, setting $P_{u}\!=\!P$, \emph{i.e.}, treating the interference as noise, achieves within 1 bit of the
 sum-capacity of the channel.
 \end{theorem}
 \begin{IEEEproof}
For the IC, \emph{i.e.}, $C_{12}=0$, \cite{EtkinIT07} proves the
theorem. The achievable sum-rate becomes:
 \begin{equation}\label{eq: ach_sum_nl}
    R_{\text{sum}}^{\text{nl}}=\log(1+\frac{P}{1+aP}).
 \end{equation}
To extend the result of \cite{EtkinIT07}, to $C_{12}>0$ case, this rate is compared to the upper bounds obtained from
CgRC.

For $a\leq 1$, the difference $\Delta_{a\leq 1}^{\text{nl}}$ between the achievable
sum-rate of Eq. (\ref{eq: ach_sum_nl}) and the upper bound (\ref{eq: ub_1}) becomes:
\begin{IEEEeqnarray}{rl}
\Delta_{a \leq 1}^{\text{nl}}&=\frac{1}{2}\bigg[\log\big(1+(1+\sqrt{a})^{2}P\big)+\log(\frac{1+P}{1+aP})-\nonumber\\
{}&\quad \quad -2\log\big(\frac{1+aP+P}{1+aP}\big)\bigg]\nonumber\\
{}&=\frac{1}{2}\bigg[\log\big(\frac{1+(1+\sqrt{a})^{2}P}{1+aP+P}\big)+\log(\frac{(1+P)(1+aP)}{1+aP+P})\bigg]\nonumber\\
{}&=\frac{1}{2}\bigg[\log\big(1+\frac{2\sqrt{a}P}{1+aP+P}\big)+\log\big(1+\frac{aP^{2}}{1+aP+P}\big)\bigg]\nonumber\\
{}&\stackrel{\text{(a)}}{\leq}\frac{1}{2}\bigg[\log\big(1+\frac{aP+P}{1+aP+P}\big)+\log\big(1+\frac{P}{1+aP+P}\big)\bigg]\nonumber\\
{}&\leq 1,\nonumber
\end{IEEEeqnarray}
where (a) follows from $2\sqrt{a}\leq (1+a)$ and $aP\leq 1$.

Similarly, by comparing the achievable sum-rate of Eq. (\ref{eq: ach_sum_nl}) and the second term in
the upper bound of Eq. (\ref{eq: ub_3}), it can be shown that the gap is
not greater than 1 bit for $a>1$ as follows:
\begin{align*}
\Delta_{a> 1}^{\text{nl}}&=\frac{1}{2}\log\big(1+(1+\sqrt{a})^{2}P\big)-\log\big(1+\frac{P}{1+aP}\big)\nonumber\\
{}&=\frac{1}{2}\Big[\log\big(\frac{1+(1+\sqrt{a})^{2}P}{1+aP+P}\big)+\log\big(\frac{(1+aP)^{2}}{1+aP+P}\big)\Big]\nonumber\\
{}&=\frac{1}{2}\Big[\log\big(1\!+\!\frac{2\sqrt{a}P}{1\!+\!aP\!+\!P}\big)+\log\big(1\!+\!\frac{(aP)^{2}\!+\!aP\!-\!P}{1\!+\!aP\!+\!P}\big)\Big]\nonumber\\
{}&\stackrel{\text{(a)}}{\leq}\frac{1}{2}\bigg[\log\big(1+\frac{aP+P}{1+aP+P}\big)+\log\big(1+\frac{1+aP-P}{1+aP+P}\big)\bigg]\nonumber\\
{}&\leq \frac{1}{2}[1+1],\nonumber\\
{}&\leq 1,\nonumber
\end{align*}
where in (a), $2\sqrt{a}$ and $(aP)^{2}$ are replaced by the larger quantities $(1+a)$ and $1$, respectively.

 \end{IEEEproof}

\section{Conclusion}
We achieved within two bits of the sum-capacity of the symmetric interference channel with unidirectional cooperation by applying simple HK scheme in conjunction with cooperative communication. In particular, we proposed a power allocation strategy to divide power between private, common and cooperative codewords for each transmitter in the weak interference regime. For the strong interference regime we simplified the achievablity scheme to use one or both of common and cooperative codebooks for different scenarios. The simplified schemes ensure the maximum gap of 1 bit from the sum-capacity of the channel.
\appendices
\section{Proof of Theorem 1}\label{appendix: Proof_1}
The proof consists of the following steps:
\begin{enumerate}
\item $R_{1}-R_{i}\leq\frac{1}{2}$ for $i\in\{3,4,5\}$,\\
\item $\Delta_{1}\triangleq R_{\text{ub}}^{(1)}-R'_{1} \leq 1.2$,\\
\item $\Delta_{2}\triangleq R_{\text{ub}}^{(2)}-R'_{2} \leq 2$,\\
\end{enumerate}
where $R_{1},\cdots, R_{5}$ are defined in Eq. (\ref{eqnarr: R1_5 }), and
\begin{equation*}
    R'_{i}\triangleq R_{i}+2C(\frac{P_{u}}{2}) \quad \text{for} \quad i\in \{1,2\}.
\end{equation*}
To prove the first step, it is sufficient to show
\begin{equation*}
    \Delta'\triangleq C(\frac{(2+a)P_{w}}{P_{u}+2})-C(\frac{(1+a)P_{w}}{P_{u}+2})\leq \frac{1}{2},
\end{equation*}
which is true since:
\begin{equation*}
   \Delta'=\frac{1}{2}\log(1+\frac{P_{w}}{P_{w}(1+a)+P_{u}+2})\leq \frac{1}{2},
\end{equation*}
where $b$ is defined in Eq. (\ref{eq: b}).
 $R'_{1}$ and $R'_{2}$ can be written as:
\begin{IEEEeqnarray*}{ll}
   R'_{1}&=\frac{1}{2}\log\Big(\frac{(2a+1)\big[ab(1+(2+a)P)+a-2\sqrt{a}\big]}{4a^{2}(1+b)}\Big),\\
   R'_{2}&=C_{12}+\log\Big(\frac{(2a+1)+b(1+a+a^{2}P)}{a(1+b)}\Big).
   \end{IEEEeqnarray*}
We define:
\begin{IEEEeqnarray*}{ll}
X&\triangleq abP \\
A&\triangleq 1+b \\
B&\triangleq aA \\
C&\triangleq (2+a)(2a+1) \\
D&\triangleq (2a+1)(ab+a-2\sqrt{a}).
\end{IEEEeqnarray*}
Hence:
\begin{IEEEeqnarray*}{ll}
   \Delta_{1}&=1+\frac{1}{2}\big[\log(\frac{AX+B}{CX+D})+\log(\frac{aP+a}{aP+1})\big]\\
   {}&\stackrel{\textrm{(I)}}{\leq}1+\frac{1}{2}\log(\frac{AX+B}{CX+D})\\
   {}&\stackrel{\textrm{(II)}}{\leq}1+\frac{1}{2}\log(\frac{Ab+B}{Cb+D})\\
   {}&\stackrel{\textrm{(III)}}\leq 1.2
\end{IEEEeqnarray*}

Since $a\!\leq\!1$ (I) is true. Because $AD-BC\!\leq\!0$, $\frac{AX+B}{CX+D}$ is a monotonically decreasing function of $X$ for $\frac{-D}{C}\!\leq\!X$. Noting that $aP\!\geq\!1$, it is straightforward to show $\frac{-D}{C}\!\leq\!b\!\leq\!X$, therefore (II) is also true. As $\frac{Ab+B}{Cb+D}$ is only a function of $a$, (III) can be simply verified.
In addition:
\begin{equation*}
   \Delta_{2}=1+\log\Big(\frac{a(b+1)\big[P+(aP+1)^{2}\big]}{(aP+1)\big[b(a+1+a^{2}P)+2a+1\big]}\Big),
\end{equation*}
and it is easy to show that $\Delta_{2}\leq 2$. Therefore, the gap is at most 2 bits which occurs when $R=R_{2}$, where $R$ is defined in Eq. (\ref{eq: R_min}).

\end{document}